\documentclass[%
 reprint,
superscriptaddress,
 amsmath,amssymb,
 aps,
prl,
]{revtex4-1}
\usepackage{multirow}
\usepackage{graphicx}
\usepackage{dcolumn}
\usepackage{bm}
\usepackage{todonotes}
\usepackage{simplewick}
\usepackage{hyperref}
\usepackage{physics}
\usepackage{verbatim}
\usepackage{float}
\usepackage{array}
\usepackage[utf8]{inputenc}
\usepackage[T1]{fontenc}
\hypersetup{
     colorlinks,
     linkcolor={blue!50!black},
     citecolor={blue!50!black},
     urlcolor={blue!80!black}
}
\graphicspath{{./figures/}}

\usepackage{xspace}
\makeatletter
\DeclareRobustCommand\onedot{\futurelet\@let@token\@onedot}
\def\@onedot{\ifx\@let@token.\else.\null\fi\xspace}

\makeatother

\newcommand{\zbb}{0\nu\beta\beta}
\newcommand{\tbb}{2\nu\beta\beta}

\begin{document}
\title{Correlation of $0\nu\beta\beta$ decay nuclear matrix elements with nucleon-nucleon phase shifts}%

\author{A. Belley}%
\affiliation{TRIUMF 4004 Wesbrook Mall, Vancouver BC V6T 2A3, Canada}%
\affiliation{Department of Physics \& Astronomy, University of British Columbia, Vancouver, British Columbia V6T 1Z1, Canada}

\author{J. Pitcher}%
\affiliation{TRIUMF 4004 Wesbrook Mall, Vancouver BC V6T 2A3, Canada}%
\affiliation{Department of Physics \& Astronomy, University of British Columbia, Vancouver, British Columbia V6T 1Z1, Canada}
\affiliation{Department of Computer Science, University of Toronto, Toronto, Ontario, M5S 3H2, Canada}

\author{T. Miyagi}%
\affiliation{Center for Computational Sciences, University of Tsukuba, 1-1-1 Tennodai, Tsukuba 305-8577, Japan}
\affiliation{Technische Universit\"at Darmstadt, Department of Physics, 64289 Darmstadt, Germany}
\affiliation{ExtreMe Matter Institute EMMI, GSI Helmholtzzentrum f\"ur Schwerionenforschung GmbH, 64291 Darmstadt, Germany}
\affiliation{Max-Planck-Institut f\"ur Kernphysik, Saupfercheckweg 1, 69117 Heidelberg, Germany}

\author{S. R. Stroberg}%
\affiliation{Department of Physics and Astronomy, University of Notre Dame, Notre Dame, IN 46556 USA}

\author{J. D. Holt}%
\affiliation{TRIUMF 4004 Wesbrook Mall, Vancouver BC V6T 2A3, Canada}%
\affiliation{Department of Physics, McGill University, Montr\'eal, QC H3A 2T8, Canada}%

\begin{abstract}

We present an ab initio study of the correlation between nuclear matrix elements of $\zbb$ decay and nucleon-nucleon scattering phase shifts in the $^1S_0$ channel. 
Starting from thirty-four statistically weighted interactions derived from chiral effective field theory, we apply the valence-space in-medium similarity renormalization group to calculate nuclear matrix elements in  four key experimental isotopes. 
Comparing with the $^1S_0$-channel phase shifts given from each interaction, in all cases we observe a strong correlation for scattering energies above 75 MeV.  
Furthermore, a global sensitivity analysis, enabled by newly developed machine-learning emulators, confirms that the nuclear matrix elements of the decay depend primarily on the $C_{1S0}$ low-energy constant, which is associated with the phase shifts in that partial wave. 
These results provide the first clear correlation between $\zbb$ decay nuclear matrix elements and a \textit{measured} observable and will therefore serve as a crucial component in ongoing and future refinements of ab initio uncertainty estimates.

\end{abstract}

\maketitle

Neutrinoless double-beta ($\zbb$) decay is a hypothetical nuclear decay that, if observed, would greatly change our understanding of the fundamental laws of nature~\cite{Agos24RMP}. 
This decay, in which two neutrons transform into two protons via emission of two leptons (electrons) with no accompanying anti-leptons (antineutrinos), would violate the lepton-number conservation predicted by the Standard Model (SM) of particle physics.
In fact, it is the most sensitive probe of lepton-number violation currently known~\cite{Gouvea:2013}; an observation could, in turn, elucidate the puzzle of matter-antimatter asymmetry in the universe~\cite{Perez:2022}.  
Furthermore, for this decay to occur, the neutrino must be its own antiparticle (a Majorana fermion)~\cite{Schechter:1982}, unlike all other fundamental fermions, which have distinct antiparticles (Dirac fermions). 
While the simplest mechanism that can give rise to this decay is the exchange of light Majorana neutrinos, many other beyond-standard-model (BSM) mechanisms have been proposed~\cite{Engel2017, Cirigliano2022, Graf2022}. 
In principle, if this decay is observed in multiple candidate isotopes, the specific mechanisms at play can be inferred, allowing $\zbb$ decay searches to assist high-energy colliders in the discovery of new physics~\cite{Helo2013}. 
However, for this to be practical, the nuclear matrix elements (NMEs) which relate the half-life of the decay to underlying new physics, and can only be obtained through nuclear theory, must be known with sufficient accuracy~\cite{Cirigliano2022, Graf2022}.

\begin{figure*}[th]
    \centering
    \includegraphics[width=\linewidth]{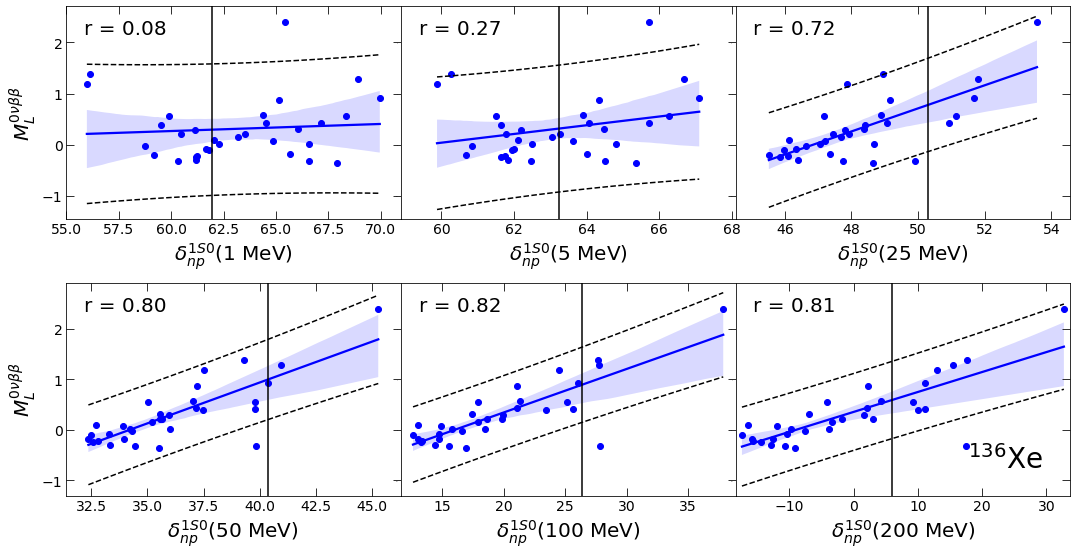}
    \caption{ Correlation between the neutron-proton phase shift in the $^1$S$_0$ partial wave at 1, 5, 25, 50, 100, and 200 MeV lab energies and the long-range NMEs of $\zbb$ decay in the key candidate isotope $^{136}$Xe, using 34 non-implausible samples of the LECs of a delta-full chiral EFT interaction at NNLO from Ref.~\cite{Hu22Pb208}. The vertical line shows the extracted experimental phase shifts taken from the Granada analysis~\cite{Perez2013}. 
    }
    \label{fig:correlation}
\end{figure*}

Reliable calculations of the NMEs have proven to be a significant challenge to the field of nuclear theory for decades. 
Nuclear models, constrained to reproduce selected experimental data, generally show a spread in results~\cite{Engel2017} much too large to pin down a specific decay mechanism, should a discovery be made. 
This is even more problematic when considering all models rely on uncontrolled approximations and cannot rigorously quantify their respective uncertainties. 
One emerging strategy to constrain NMEs has been to explore possible correlations with nuclear observables, either by varying phenomenological interactions in the same nuclei~\cite{Horoi:2022, Horoi:2023}, or relating observables in multiple nuclei across the nuclear chart~\cite{Shimizu:2018, Brase2022}. 
Within various models, strong correlations have been found between $\zbb$-decay NMEs  and $\tbb$ decay~\cite{Horoi:2022,Horoi:2023,Jokiniemi:2023,Jokiniemi:2023-2vbb}, double Gamow-Teller (DGT) transitions~\cite{Shimizu:2018, Brase2022, Jokiniemi:2023}, and double-magnetic-dipole decays~\cite{Jokiniemi:2023}.

In contrast, recent advances in first-principles, or ab initio, nuclear theory~\cite{Hergert2020,Ekstrom2023} now allow for robust statistical uncertainty quantification for calculated nuclear observables, even to heavy systems~\cite{Hu22Pb208}. 
This is feasible since ab initio methods rely on systematically improvable approximations to solve the nuclear many-body problem, starting directly from the inter-nucleon forces derived from chiral effective field theory (EFT)~\cite{Epelbaum2009,Machleidt2011}. 
This is particularly exciting for BSM physics searches, where nuclear theory inputs are required to interpret current experimental limits in terms of excluded new physics (see, e.g., Refs.~\cite{Brodeur:2023,Hu22SDDM,Joki23OMC,Arrow24}) and is crucial for the specific case of $\zbb$ decay.
Ab initio methods, which reproduce single-beta decays without quenching of $g_A$~\cite{Gysbers2019} and have been benchmarked against quasi-exact calculations in light nuclei~\cite{Yao2021}, have now computed first NMEs for essentially all experimentally relevant $\zbb$-decay isotopes~\cite{Yao2020,Belley2021,Novario2021,Belley2023_texe}.
While a more limited spread in NMEs is seen, compared to that from nuclear models, a true uncertainty quantification was lacking, until a very recent first analysis~\cite{Belley:2024}. 
In addition, ab initio methods have investigated the correlation between the DGT transitions and $\zbb$ NMEs, where the correlation between different isotopes is weaker than that seen in nuclear models, the correlation obtained by varying the nuclear interaction was the strongest found at the time~\cite{Yao:2022-DGT}.
This highlights the importance of reliable ab initio calculations as well as experimental efforts searching for DGT transitions. 
In this Letter, we expand these efforts by demonstrating the correlation of $\zbb$ decay NMEs, in multiple candidate isotopes, with an \textit{observed} nuclear property:~nucleon-nucleon scattering phase shifts in the $^{1}S_{0}$ partial wave. 
In particular, we focus on the long-range part of the standard light-neutrino exchange, since it dominates the decay, and further present a simple physical explanation for why this term should be correlated with the phase shifts.

Our calculations start from chiral EFT~\cite{Epelbaum2009,Machleidt2011}, which provides a systematic low-energy expansion of quantum chromodynamics (QCD) in terms of nucleon-nucleon (NN) and three-nucleon (3N) interactions, as well as a consistent prescription of electroweak forces. 
In this framework, long-range nuclear forces are mediated through pion exchanges, while short-distance physics is encoded in low-energy constants (LECs), typically constrained to reproduce few-body data. 
In particular, we use a formulation with an explicit $\Delta$ isobar degree of freedom at next-to-next-to leading order (NNLO), as presented in Ref.~\cite{Ekstrom2018,Jiang2020}, in which there are 17 unconstrained LECs. 
Here, we take the 34 statistically weighted interactions constructed in Ref.~\cite{Hu22Pb208} as a representative sample of the non-implausible parameter space, which were obtained from of $\sim 10^{9}$ parameter sets via history matching, based on their prediction of selected observables in light nuclei.

To ultimately assign an uncertainty to our theoretical calculations, it is crucial to study the dependence of a given observable on the LECs. 
This is achieved with the sampling/importance resampling (SIR) technique~\cite{Smith1992}, which makes uses of Bayes' theorem to assign weights to parameter samples via a likelihood function representing how well each sample performs for a set of calibration observables~\cite{Hu22Pb208}.
The implicit assumption in this approach is that parameter sets that more accurately reproduce the calibration observables should also be more accurate for the target observable.
The issue for $0\nu\beta\beta$ decay is that none of the calibration observables proposed in previous studies have been found to correlate with the NMEs~\cite{Belley:2022}. 
As mentioned above, in ab initio calculations the only correlation so far observed with $\zbb$ decay NMEs has been DGT transitions~\cite{Yao22DGT}, which are challenging experimentally, thus leaving the likelihood function essentially unconstrained. 
We now, however, discuss a correlation between nucleon-nucleon phase shifts in the $^{1}S_{0}$ partial wave between 50 and 200 MeV and the $\zbb$ NME in multiple candidate isotopes, opening a pathway towards a much-needed rigorous uncertainty quantification.

To approximately solve the many-body problem required to obtain the NMEs, we use the valence-space formulation of the in medium similarity renormalization group (VS-IMSRG) \cite{Stroberg2017,Stro19ARNPS,Miya20MS}. 
Starting from a harmonic oscillator basis of 15 major shells and 3N force matrix elements sufficient to achieve convergence~\cite{Miyagi2022}, we construct an approximate unitary transformation~\cite{Morr15Magnus} to decouple a valence-space Hamiltonian and consistently transformed $\zbb$ transition operator~\cite{Parz17Trans}. 
We then perform an exact diagonalization within the valence space using the KSHELL shell-model code~\cite{Shimizu2019}. 
Since this decoupling is unitary, this method would be exact if no truncation was made.
Here, we use the IMSRG(2) approximation, where all operators arising in nested commutators of the expansion are truncated at the two-body level.

\begin{figure}
    \centering
    \includegraphics[width=\columnwidth]{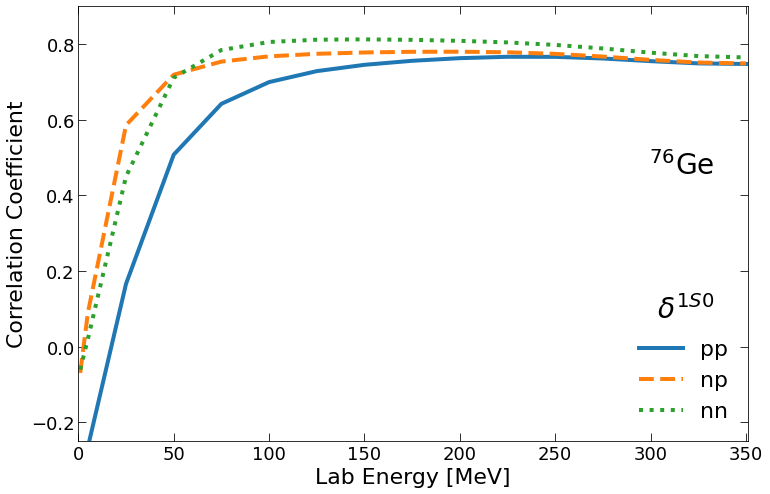}
    \includegraphics[width=\columnwidth]{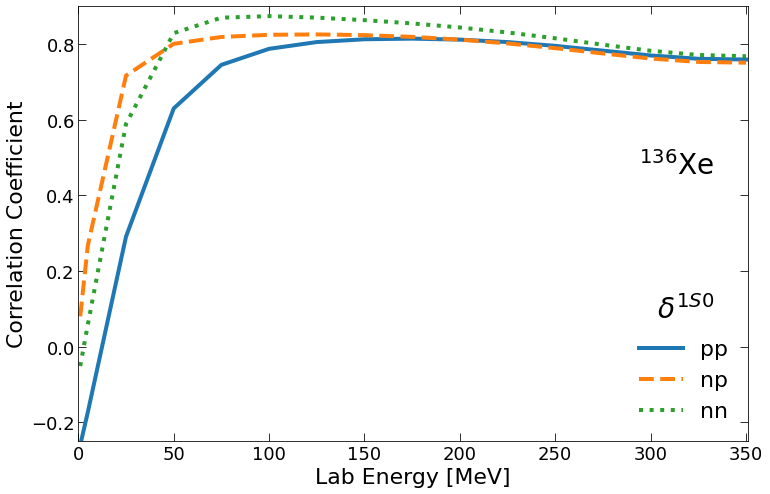}
    \caption{Correlation coefficient between the long-range NMEs of $\zbb$ decay in $^{76}$Ge and $^{136}$Xe, and the proton-proton (pp, full), neutron-proton (np, dashed) and neutron-neutron (nn, dotted) phase shift in the $^{1}S_0$ partial wave for energies up to 350 MeV. 
    We note that for all cases but the pp phase-shift in $^{76}$Ge, a strong correlation is found for lab energies $\geq 75$~MeV. 
    }
    \label{fig:coefficient_vs_energy}
\end{figure}

 To measure the correlation between the matrix elements and the phase shifts,  we use the Pearson's r coefficient, or correlation coefficient, defined as~\cite{Freedman2007} 
\begin{align}
    r = \dfrac{\sum_i (x_i -\bar{x})(y_i-\bar{y}_i)}{\sqrt{\sum_i(x_i-\bar{x})\sum_i(y_i-\bar{y})}}
\end{align}
where $x_i$ and $y_i$ are the values of the two observables for the $i$th LEC sample and $\bar{x}$ and $\bar{y}$ are the averages of each observable.
Fig.~\ref{fig:correlation} shows the correlation between the neutron-proton phase shift in the $^{1}S_0$ partial-wave channel at select lab energies between 1 MeV and 200 MeV in $^{136}$Xe. 
Correlation coefficients for additional candidate isotopes, $^{48}$Ca, $^{76}$Ge, and $^{130}$Te for the three states of the isospin triplet are presented in the supplemental material~\cite{SupplementalMaterial}. 

In all cases studied, we find a strong correlation with the np phase shift at lab energies higher than $\sim$50 MeV.
This is illustrated better in Fig.~\ref{fig:coefficient_vs_energy}, which shows the correlation coefficient as a function of the energies for the proton-proton (pp), neutron-proton (np) and neutron-neutron (nn) phase shift with the NMEs of $^{76}$Ge and $^{136}$Xe. 
While there is no apparent correlation at low energies, the onset and increase is rapid with lab energy, before reaching a maximum near 100 MeV, 125 MeV and 175 MeV for the nn, np and pp phase shifts, respectively. 
The pp phase shifts reach their maximum correlation at higher energies due to the Coulomb repulsion between the two scattering protons that needs to be overcome. 
As shown in Supplemental Material, these features are consistent across all candidate isotopes.

\begin{figure*}[t]
    \centering
    \includegraphics[width=\textwidth]{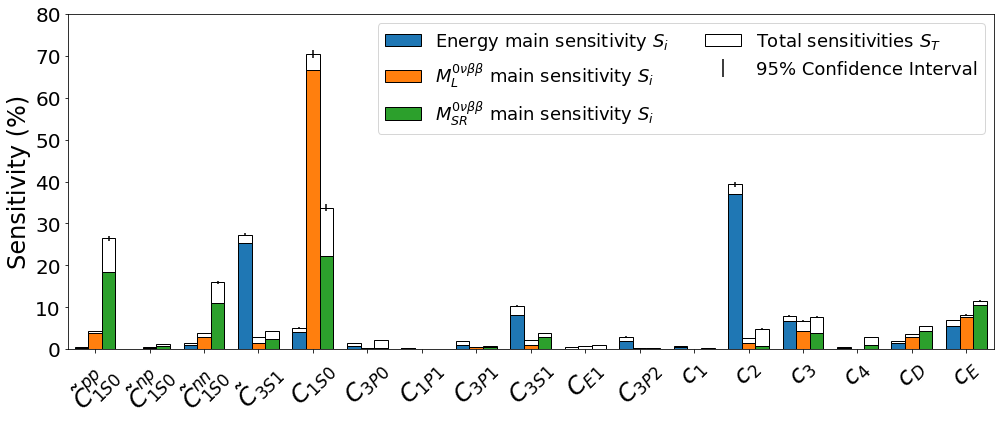}
    \caption{Sensitivity of the ground state energy (blue), the long-range NMEs $M^{0\nu\beta\beta}_L$ and the short-range matrix elements $M^{0\nu\beta\beta}_{SR}$ of $\zbb$ decay for the isotope $^{76}$Ge to the low-energy constants arising in nuclear interactions from chiral EFT.
    }
    \label{fig:sensitivity}
\end{figure*}

To better understand the origin of this correlation, we perform a global sensitivity analysis of the NMEs following the formalism of Ref.~\cite{Ekstrom2019}. 
Generally this allows us to probe the extent to which each input parameter contributes to the variance of the final result~\cite{SOBOL2001271}, i.e., in this case, how the calculated NMEs depends on the individual LECs within chiral EFT. 
This level of statistical analysis requires a tremendous number of sample calculations, far more than could realistically be carried out using full VS-IMSRG framework. 
To this end, we utilize the newly developed emulator of the VS-IMSRG, the so-called multi-fidelity, multi-output deep Gaussian process (MM-DGP) algorithm~\cite{Pitcher_inprep}, which greatly increases computational speed, allowing for approximately 8000 calculations to be performed in a few seconds after the emulator is trained.
To perform the final global sensitivity analysis, we use the SALib~\cite{Herman2017,Iwanaga2022} python package to compute the sensitivities and their bootstrapped 95\% confidence interval for each parameter.

In Fig.~\ref{fig:sensitivity} we show the results of this sensitivity analysis for the ground-state energy, as well as the long- and short-range $\zbb$ NMEs, of $^{76}$Ge. 
First, we note that the relative importance of LECs for the ground-state energy qualitatively agrees with results of Ref.~\cite{Ekstrom2019}.
While an exact comparison is not possible, as the interactions used in that study did not include explicit delta degrees of freedom, this is nevertheless a reasonable verification that our MM-DGP emulator is reliable and consistent with analogous analyses. 
We then see that the long-range NMEs are almost entirely dominated by contributions coming from the $C_{1S0}$ LEC, which is associated with the strength of the short/medium-range nuclear interaction in the $^1S_0$ partial wave. 
The contribution mostly comes from the first-order effect $S_i$, reinforcing that this LEC is dominant, rather than the NMEs being depending on some combination of LECs. 
Since this LEC also determines the phase shifts in this partial wave, it is not surprising that a strong correlation between the two quantities is observed. 
When performing the same exercise for the short-range NMEs, however, there is a clear dependence on a number of LECs, also shown in Fig.~\ref{fig:sensitivity}, thus making the correlation weaker for this term. 
This is because the coupling constant of the short-range contact term~\cite{Ciri18contact,Ciri21contact} depends on $\tilde{C}^{pp}_{1S0}$ and $\tilde{C}^{nn}_{1S0}$ while the short-range NME themselves depends on $C_{1S0}$.
Regardless, as the long-range term is dominant, this correlation still allows us to now strongly constrain the NMEs of $\zbb$ decay.

Finally, at NNLO, only two LECs are active in the $^1S_0$ partial wave: $C_{1S0}$, and $\tilde{C}^{nn}_{1S0}$ (in addition to the pion-nucleon coupling constants, which are already tightly constrained~\cite{Hoferichter2016}).
The leading-order ($Q^0$) contact $\tilde{C}^{nn}_{1S0}$ is required to match the scattering length, so the only freedom at higher energies comes from the NLO ($Q^2$) contact $C_{1S0}$. 
Hence the reason that all phase shifts above $\sim 50$ MeV show similar correlation with the NME:~they are all controlled by the same LEC.
Of course, additional LECs appear in the $^1S_0$ partial wave at higher orders in the EFT,  multiplying operators with higher powers of momentum; these will have greater impact at higher scattering energies. 
If the correlation were driven by the physics at higher energies, it would not be useful due to sensitivity to unconstrained higher-order LECs. 
However, it is reasonable to expect that the relevant scale is dictated by the fermi momentum in nuclei and the momentum transfer of the decay, corresponding to $\lesssim 100$~MeV lab frame kinetic energy. 
To confirm this expectation, we have included the N$^3$LO ($Q^3$) contact $D_{1S0}$, to allow separate variation of medium- and high-energy phase shifts (see supplemental material~\cite{SupplementalMaterial}). 
We indeed find that the correlation is driven by the medium-energy range, and so should persist in higher-order calculations.

We further note that an even stronger correlation can be found between the long-range NME and the long-range two-body amplitudes (described in Ref.~\cite{Cirigliano2021}) of the decay, with the momentum of both particles set at 200 MeV/c. 
This correlation, however, is not seen with the total amplitudes, as the short-range component contributes more strongly to the amplitudes than the NMEs. 
While lattice QCD calculations would, in principle, obtain the full amplitude, they will not completely constrain the value of the NMEs on their own. 
Instead, they would, quite importantly, permit us to greatly constrain the number of LEC samples that give reasonable amplitudes. 
By combining this constraint with the correlation with the phase-shift, the allowed space for the LECs would be greatly reduced.

In conclusion, we find that the long-range $\zbb$ decay NMEs are strongly correlated with the \textit{observed} phase shifts of the isospin triplet in the $^1S_0$ partial wave at energies $\geq$ 75 MeV. 
By performing a global sensitivity analysis of the nuclear matrix elements, we find that the $C_{1S0}$ LEC has the largest contribution to the NMEs, which explains the correlation. 
This correlation allows us to quantify the uncertainty from the input forces for the NMEs of $\zbb$ decay, as was done in Ref.~\cite{Belley:2024}.
We note that the present study does not account for uncertainties related to approximations made in solving the many-body problem; this must also be studied carefully.
Due to this, the correlation cannot be used to extract the value of the matrix elements directly.
Nonetheless, this correlation and more generally the development of emulators to bypass the computational cost of  many-body methods finally open a way towards theoretical nuclear calculation with a reliable uncertainty. This is crucial in the case of $\zbb$ decay in order to identify the mechanism at play in the decay, but also for all other searches of new physics that rely on nuclear theory inputs to constrain extensions of the Standard Model. 

\begin{acknowledgments}

We thank A. Ekström, C. Forssén, G. Hagen, and W. G. Jiang for providing the interaction samples used in this work and L. Jokiniemi and A. Schwenk for insightful discussions.
The IMSRG code used in this work makes use of the Armadillo \texttt{C++} library \cite{Sanderson2016, Sanderson2018}.
TRIUMF receives funding via a contribution through the National Research Council of Canada. 
This  work was further supported by NSERC under grants SAPIN-2018-00027 and RGPAS-2018-522453, the Arthur B. McDonald Canadian Astroparticle Physics Research Institute, the Canadian Institute for Nuclear Physics, the US Department of Energy (DOE) under contract DE-FG02-97ER41014, the European Research Council (ERC) under the European Union’s Horizon 2020 research and innovation programme (Grant Agreement No. 101020842), the Deutsche Forschungsgemeinschaft (DFG, German Research Foundation) -- Project-ID 279384907 -- SFB 1245, and JST ERATO Grant No. JPMJER2304, Japan..
Computations were performed with an allocation of computing resources on Cedar at WestGrid and The Digital Research Alliance of Canada.

\end{acknowledgments}

\bibliographystyle{apsrev4-1}
\bibliography{library}{}

\section{}
\newpage
\section{Supplemental Material}

\begin{table*}[h!]
\centering
\begin{tabular}{cc|cccccccccccccccc}
\multicolumn{1}{l}{} &
  \textbf{} &
  \multicolumn{16}{c}{\textbf{Pearson's r coefficient}} \\
  \hline
\multicolumn{2}{c|}{\textbf{Energy (MeV)}} &
  \textbf{1} &
  \textbf{5} &
  \textbf{25} &
  \textbf{50} &
  \textbf{75} &
  \textbf{100} &
  \textbf{125} &
  \textbf{150} &
  \textbf{175} &
  \textbf{200} &
  \textbf{225} &
  \textbf{250} &
  \textbf{275} &
  \textbf{300} &
  \textbf{325} &
  \textbf{350} \\
  \hline
\multirow{3}{*}{\textbf{$^{48}$Ca}} &
  \textbf{pp} &
  -0.213 &
  -0.117 &
  0.373 &
  0.708 &
  0.810 &
  0.844 &
  0.856 &
  0.862 &
  0.864 &
  0.863 &
  0.860 &
  0.852 &
  0.840 &
  0.827 &
  0.817 &
  0.814 \\
 &
  \textbf{np} &
  -0.087 &
  0.118 &
  0.665 &
  0.789 &
  0.820 &
  0.832 &
  0.838 &
  0.841 &
  0.842 &
  0.841 &
  0.837 &
  0.830 &
  0.819 &
  0.808 &
  0.799 &
  0.797 \\
 &
  \textbf{nn} &
  -0.117 &
  -0.018 &
  0.520 &
  0.802 &
  0.863 &
  0.877 &
  0.879 &
  0.878 &
  0.875 &
  0.871 &
  0.865 &
  0.856 &
  0.843 &
  0.829 &
  0.819 &
  0.815 \\
  \hline
\multirow{3}{*}{\textbf{$^{76}$Ge}} &
  \textbf{pp} &
  -0.341 &
  -0.265 &
  0.166 &
  0.508 &
  0.642 &
  0.700 &
  0.729 &
  0.746 &
  0.756 &
  0.763 &
  0.767 &
  0.767 &
  0.762 &
  0.755 &
  0.749 &
  0.748 \\
 &
  \textbf{np} &
  -0.069 &
  0.090 &
  0.586 &
  0.719 &
  0.754 &
  0.768 &
  0.775 &
  0.778 &
  0.780 &
  0.780 &
  0.779 &
  0.775 &
  0.767 &
  0.759 &
  0.752 &
  0.749 \\
 &
  \textbf{nn} &
  -0.060 &
  0.020 &
  0.450 &
  0.711 &
  0.785 &
  0.806 &
  0.812 &
  0.813 &
  0.812 &
  0.809 &
  0.805 &
  0.798 &
  0.789 &
  0.778 &
  0.769 &
  0.765 \\
  \hline
\multirow{3}{*}{\textbf{$^{130}$Te}} &
  \textbf{pp} &
  -0.258 &
  -0.153 &
  0.378 &
  0.750 &
  0.869 &
  0.911 &
  0.930 &
  0.932 &
  0.931 &
  0.924 &
  0.913 &
  0.896 &
  0.876 &
  0.857 &
  0.845 &
  0.842 \\
 &
  \textbf{np} &
  0.033 &
  0.249 &
  0.788 &
  0.894 &
  0.918 &
  0.927 &
  0.929 &
  0.928 &
  0.923 &
  0.914 &
  0.901 &
  0.885 &
  0.866 &
  0.848 &
  0.836 &
  0.833 \\
 &
  \textbf{nn} &
  -0.172 &
  -0.065 &
  0.518 &
  0.841 &
  0.919 &
  0.941 &
  0.947 &
  0.946 &
  0.941 &
  0.932 &
  0.918 &
  0.900 &
  0.879 &
  0.859 &
  0.845 &
  0.842 \\
  \hline
\multirow{3}{*}{\textbf{$^{136}$Xe}} &
  \textbf{pp} &
  -0.260 &
  -0.172 &
  0.291 &
  0.630 &
  0.745 &
  0.788 &
  0.806 &
  0.813 &
  0.815 &
  0.812 &
  0.806 &
  0.796 &
  0.783 &
  0.770 &
  0.762 &
  0.760 \\
 &
  \textbf{np} &
  0.080 &
  0.267 &
  0.717 &
  0.801 &
  0.819 &
  0.825 &
  0.826 &
  0.824 &
  0.819 &
  0.812 &
  0.802 &
  0.790 &
  0.776 &
  0.762 &
  0.753 &
  0.751 \\
 &
  \textbf{nn} &
  -0.510 &
  0.059 &
  0.589 &
  0.829 &
  0.870 &
  0.874 &
  0.870 &
  0.864 &
  0.855 &
  0.844 &
  0.831 &
  0.816 &
  0.799 &
  0.783 &
  0.772 &
  0.768\\
  \hline
\end{tabular}
\caption{Correlation coefficient between the long-range NMEs of $\zbb$ in 4 candidate isotopes and the proton-proton  (pp), neutron-proton (np) and neutron-neutron (nn) phase shift in the $^{1}$S$_0$ partial wave with energies going from 1 MeV to 350 MeV}
\label{tab:r_coeffecient}
\end{table*}

\begin{figure*}[h!]
    \centering
    \includegraphics[width=\linewidth]{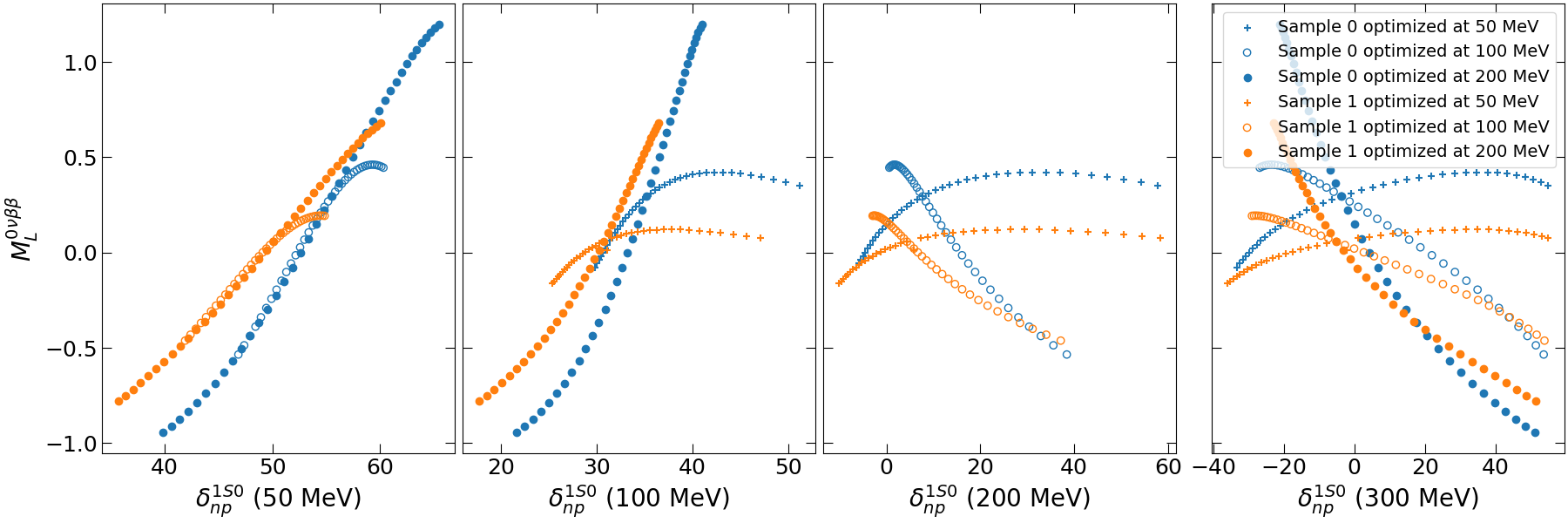}
    \caption{Correlation between the neutron-proton phase-shift in the $^1$S$_0$ partial wave at 50, 100, 200 and 300 MeV and the long-range NMEs of $\zbb$ in the candidate isotope  $^{48}$Ca using 2 non-implausible samples of the LECs of a delta-full $\chi$-EFT interaction at NNLO and adding the $D_{1S0}$ term of the N3LO interaction. The potential in the 1S0 partial wave is therefore given by $V(p,p') = \tilde{C}_{1S0} + C_{1S0}(p^2+p'^2)+D_{1S0}p'^2p^2 + V_{1\pi}(p,p') + V_{2\pi}(p,p')$ with a regulator function. Note  that $\tilde{C}_{1S0}$ is different for the nn, np and pp cases due to  isospin breaking and that the full N3LO potential contains another term given by $\hat{D}_{1S0}(p'^4+p^4)$ which we neglected here for simplicity.  The values of $C_{1S0}$ are adjusted to fix the phase-shift at either 50, 100 or 200 MeV. For clarity, we do not show the results with fixed value of the phase-shift. We find that fixing the phase-shift at higher energy, which forcefully breaks the correlation at this energy, has little effect on the correlation at lower energies. However, the opposite is not true as breaking the correlation at low energy further breaks it at higher energies. As the N3LO corrections mostly affect higher energies in the phase-shift, the inclusion of the higher orders may break the correlation at these energies but should leave the correlation around 50-100 MeV mostly unaffected.
    }
    \label{fig:correlation_Dopt}
\end{figure*}

\end{document}